\documentclass[twocolumn,pra,amsmath,amssymb,unsortedaddress,superscriptaddress,showpacs]{revtex4}

\usepackage{latexsym,epsfig,graphicx}
\usepackage{dcolumn}
\usepackage{bm}
\usepackage{color} 

\begin{document}

\title{Probing condensate order in deep optical lattices}

\affiliation{University of Illinois at Urbana--Champaign, Urbana,
Illinois 61801, USA} \affiliation{Wellesley College, Wellesley,
Massachusetts 02481, USA}

\author{Kuei Sun}
\affiliation{University of Illinois at Urbana--Champaign, Urbana,
Illinois 61801, USA}
\author{Courtney Lannert}
\affiliation{Wellesley College, Wellesley, Massachusetts 02481,
USA}
\author{Smitha
Vishveshwara} \affiliation{University of Illinois at
Urbana--Champaign, Urbana, Illinois 61801, USA}
\date{\today}


\begin{abstract}
We study interacting bosons in optical lattices in the
weak-tunneling regime in systems that exhibit the coexistence of
Mott-insulating and condensed phases. We discuss the nature of the
condensed ground state in this regime and the validity of the
mean-field treatment thereof. We suggest two experimental
signatures of condensate order in the system. (1) We analyze the
hyperfine configuration of the system and propose a set of
experimental parameters for observing radio-frequency spectra that
would demonstrate the existence of the condensed phase between
Mott-insulating phases. We derive the structure of the signal from
the condensate in a typical trapped system, taking into account
Goldstone excitations, and discuss its evolution as a function of
temperature. (2) We study matter-wave interference patterns
displayed by the system upon release from all confining
potentials. We show that as the density profiles evolve very
differently for the Mott-insulating phase and the condensed phase,
they can be distinguished from one another when the two phases
coexist.
\end{abstract}

\maketitle

\section{Introduction}\label{sec:I}

Ultracold bosons in optical lattices allow extensive studies of
quantum phases of condensed matter in controlled
environments~\cite{Greiner03,Bloch08}. When the tunneling energy,
relative to the interaction energy, decreases across a critical
value, the system undergoes a transition from the superfluid phase
to the Mott-insulating phase~\cite{Fisher89}. This transition has
been observed in experiments by imaging matter wave interference
patterns of the system~\cite{Greiner02}. In realistic systems, a
confining trap renders the density of bosons non-uniform and, in a
sufficiently deep optical lattice (where the lattice depth is over
20 times the recoil energy), this inhomogeneous system is
predicted to have a multiple-layer structure in which
Mott-insulator layers having different occupation number are
separated by relatively thin condensate
layers~\cite{Jaksch98,DeMarco05}. The study of such inter-layer
condensates is important for a variety of reasons. From the
perspective of quantum information
schemes~\cite{Rabl03,Jaksch05,Popp06}, robust, high-fidelity
Mott-insulator states are required and the presence of a
condensate is in fact a hindrance. The condensate is also of
interest in and of itself; its ground-state properties and
excitation spectrum have been studied in various ways
~\cite{Oosten01,Taylor03,Oosten05,Sengupta05,Pupillo06a,Barankov07,Menotti08,Mitra08}
and show vast differences from more commonly encountered
condensates in free space or in the strong-tunneling limit in
shallow optical lattices. Finally, the interlayer structure is of
interest as a model realization of spatially coexisting quantum
phases separated by critical crossover regimes.

The purpose of this work is twofold: to provide a comprehensive
characterization of condensates in the vicinity of Mott-insulators
and to propose settings and measurements which would pinpoint
direct experimental signatures of the presence of condensate
interlayers in the inhomogeneous Mott-insulator-condensate shell
structure. We begin by deriving key properties of the interlayer
condensate, such as off-diagonal elements of the single-particle
density matrix, number fluctuations, and condensate fraction, in
the deep-lattice (weak-tunneling) regime. We describe the
condensate ground state with a fixed-number wave function, which
displays entanglement and off-diagonal long-range order, and
conserves the total number of particles as any exact eigenstate of
the Bose-Hubbard Hamiltonian should do. We compare this
fixed-number entangled wave function with a commonly-used
mean-field wave function. By calculating the condensate properties
in each description, we identify the conditions under which the
mean-field approximation is valid. These studies shed light on the
range of validity of mean-field descriptions of the condensate in
the deep-lattice regime. We analyze the nature of the excitations
of these condensates, including the spectrum of Goldstone mode
excitations, that would be relevant in the oft-encountered
experimental situation of two species of bosons.

Recent experiments using hyperfine transitions driven by external
radio-frequency (rf) drives have conclusively shown the
existence~\cite{Campbell06, Hazzard07} and
formation~\cite{Folling06} of a Mott-insulating shell structure in
systems where the condensate interlayers are predicted to be much
thinner than the Mott-insulating shells. Currently, however, no
strong experimental evidence unequivocally verifies the existence
or properties of the putative condensate layers between the
Mott-insulator layers~\cite{Mitra08}. In this work, we propose a
specific set of parameters for an rf experiment in which the
system is prepared and resolved as in Ref.~\cite{Campbell06}. We
find that the presence of the condensate interlayer can lead to a
two peak structure, in contrast to a single peak structure
associated with each adjoining Mott-insulator phase. We discuss
how this signature rf spectrum profile ought to be robust against
Goldstone-mode perturbations associated with spontaneous symmetry
breaking in the condensate and against low-temperature effects,
and would be sensitive to the destruction of condensate order.
This sensitivity to condensate order should allow experiments to
distinguish whether the incommensurate-density layers between
Mott-insulator shells are in the superfluid or normal fluid phase.

Another commonly used probe which has provided a wealth of
valuable information on the Mott-insulator-superfluid transition
is matter wave interference~\cite{Greiner02}. Here, we tailor the
analysis of matter-wave interference to the specific case of the
interlayer setting. We make transparent the contribution of
various terms in the fixed-number entangled wave function to
interference patterns that do not occur in the Mott-insulator
state. While discerning the condensate amidst Mott-insulator
phases through interference experiments can be a challenging task,
our analysis provides methods to do so.

The paper is outlined as follows. In Sec.~\ref{sec:II}, we review
the Bose-Hubbard model and its parameters and detail two
descriptions of the condensate ground state in the deep-lattice
regime. We review how results for a uniform system can be applied
to an inhomogeneous system using the local density approximation.
In Secs.~\ref{sec:III} and ~\ref{sec:IV}, we propose feasible
settings for experimental confirmation of the condensate layers in
the inhomogeneous system at low temperatures using rf spectroscopy
and matter-wave interference, respectively. In Sec.~\ref{sec:V},
we summarize our results.

\section{Condensed bosons in deep optical lattices}\label{sec:II}

A system of bosons in a deep optical lattice is well described by the
Bose-Hubbard Hamiltonian~\cite{Fisher89,Jaksch98,Sachdev99}:
\begin{eqnarray}
\hat H_{BH}  =  &-& J\sum_{\left\langle {ij} \right\rangle } {\left(
{\hat a_i ^\dag  \hat a_j  + \hat a_j ^\dag  \hat a_i }
\right)} \nonumber\\
&+& \sum_i {\left[ {\frac{U}{2}\hat n_i (\hat n_i - 1) - ( \mu -V_i)
\hat n_i } \right]}. \label{eqn:BHH}
\end{eqnarray}
Here $\hat a_i$ and $\hat a_i ^\dag$  are the boson creation and
annihilation operators on the $i$th lattice site and $\hat
n_i=\hat a_i ^\dag  \hat a_i$ is the number operator on the $i$th
lattice site. $\sum {_{i} }$ denotes the sum over all single sites
and $\sum {_{\left\langle {ij} \right\rangle} }$ denotes the sum
over all nearest-neighbor pairs of sites. $\mu$ is the chemical
potential, which is determined by the condition that the quantity
$\sum {_i} {\left\langle {\hat n_i } \right\rangle }$ is equal to
the total number of bosons in the system. $V_i$ is the value of
the external confining potential on site $i$. For a homogeneous
system, $V_i$ is set to zero for convenience. $J$ is the tunneling
strength, which quantifies the ability of a boson to hop from one
site to its adjacent sites. Assuming the number of particles per
site is not too large ($ \left\langle {\hat n_i } \right\rangle
\approx  {O}(1)$)~\cite{Jaksch05}, $J = - \int {d^3 {\bf{r}}w^*
({\bf{r}} - {\bf{r}}_i )} \left[ { - \hbar ^2 \nabla ^2 /2m +
V_{\rm lat} ({\bf{r}})} \right]w({\bf{r}} - {\bf{r}}_j )$, where $
{V_{\rm lat} ({\bf{r}})}$ is the lattice potential, $w({\bf{r}} -
{\bf{r}}_i )$ and $w({\bf{r}} - {\bf{r}}_i )$ are the
single-particle Wannier wave functions of the lowest Bloch band
localized to the nearest-neighbor sites $i$ and $j$, and $m$ is a
single boson's mass. $U$ is the interaction between two bosons on
a single site. $U = (4\pi \hbar ^2 a_s/m)\int {d^3 {\bf{r}}}
\left| {w({\bf{r}})} \right|^4$, where $a_s$ is the scattering
length of bosons, which is required to be small compared to the
lattice spacing for the validity of this model.

In experiments that do not use a Feshbach resonance to manipulate
the interaction between bosons, the tunable parameters are the
lattice depth, $V_0$, and the lattice spacing, $d_s$ (half of the
lattice laser's wavelength $\lambda$), which in deep lattice are
related to $J$ and $U$ by $ J = (4/\sqrt \pi )E_R (V_0 /E_R
)^{3/4} \exp [ - 2\sqrt {V_0 /E_R } ]$ and $ U = \sqrt {8/\pi }
k_R a_s E_R (V_0 /E_R )^{3/4} $, where $k_R=2\pi/\lambda$ and the
recoil energy $E_R = \hbar^2 k_R^2/(2m)$~\cite{Zwerger03}.

In Secs. IIA--IIC, we discuss the condensate state in deep optical
lattices. We discuss ground-state properties of the condensate
wave function which, as a result of its vicinity to Mott-insulator
states, is distinctly different from the condensate in the
large-tunneling limit. We begin by constructing a ground state for
a homogeneous condensate in a fixed-number truncated basis,
composed of superpositions of number states associated with the
two neighboring Mott-insulating states. Using this ground state,
we find expressions for the boson density and the number
fluctuations on each site, calculate the single-particle density
matrix, and hence obtain the condensate fraction. We find that the
maximum condensate fraction in the deep-lattice regime is
significantly smaller than one. We then consider a commonly
employed mean-field approximation to this state and show that the
single-site number fluctuations, density, and condensate fraction
of this state are identical to those in the fixed-number state. We
show that in describing the weak-tunneling condensate, using a
truncated basis of occupation number is valid, i.e., that the
error induced by truncation is of order ${O}(J/U)$. We conclude
with a discussion of the inhomogeneous system where Mott-insulator
and condensate phases coexist.

\subsection{Fixed-number condensate state}\label{sec:IIA}

In the deep lattice regime($ J/U \ll1$), the first term on the
right-hand side of Eq. (\ref{eqn:BHH}) can be treated
perturbatively to calculate physical quantities of interest, such
as the single-site number fluctuation and the condensate fraction,
in orders of $J/U$. The unperturbed basis is composed of all
possible states of the form $\prod{_i} {\left| {n_i }
\right\rangle }$ with $ \sum {n_i } $ equal to the total number of
particles $N$~\cite{Bach04}; $\left| {n_i } \right\rangle$ is
defined as $(n_i !)^{ - 1/2} (\hat a_i ^\dag )^{n_i } \left| {\rm
vac} \right\rangle$. For convenience we use the symbol $\left|
{\left| n \right\rangle } \right\rangle$ to denote a commensurate
state that has each site occupied by $n$ particles ($ \prod {_i}
{\left| {n } \right\rangle }$). The single-site number
fluctuation, denoted by $\Delta n_i ^2$, is defined as $ \Delta
n_i ^2 \equiv \left\langle {\hat n_i ^2 } \right\rangle -
\left\langle {\hat n_i } \right\rangle ^2 $; the condensate
fraction, denoted by $f_c$, can be obtained from the
single-particle density matrix of the system in the thermodynamic
limit ($N,M \to \infty$ with $N/M$ finite)~\cite{Leggett06}.
Elements of the density matrix are defined as $ \langle \hat a_i
^\dag \hat a_j \rangle $; the condensate fraction is the ratio of
the largest eigenvalue of the density matrix to the total number
of particles.

Given a system of $M$ sites, if $N$ is a multiple of $M$ ($N=nM$,
where $n$ is an integer), the ground state of the system is
$A\left| {\left| n \right\rangle } \right\rangle  + {O}(J/U)$,
where the normalization constant $A$ is equal to $1$ up to order
in $J/U$. For the state, $\Delta n_i ^2$ and $f_c$ are both zero.

If $N=nM+M_1$, where $n$ and $M_1$ are integers and $0<M_1<M$, the
ground state of the system is
\begin{equation}
\left| \Psi \right\rangle  = \sum_{\{ \eta\} } {\frac{{C_{\{
\eta\} } }}{{(n + 1)^{M_1 /2} }}\hat a_{\eta_1 } ^\dag  \hat
a_{\eta_2 } ^\dag \cdots \hat a_{\eta_{M_1 } } ^\dag  \left|
{\left| n \right\rangle } \right\rangle }  + {O}(J/U),
\label{eqn:CGS}
\end{equation}
where $\{ \eta\}  = \{ \eta_1 ,\eta_2 , \ldots ,\eta_{M_1 } \}$ is
a set of distinct integers chosen from $\{ 1,2, \ldots ,M\}$, and
$\sum {_{\{ \eta\} } } $ denotes the sum over all combinatorial
configurations. The leading term shown in $ \left| \Psi
\right\rangle $ lies in a sub-space spanned by all possible
product states of $M_1$ single-site states with occupation number
$n+1$ and $M-M_1$ single-site states with occupation number $n$;
in the small tunneling regime, this  truncation proves to be a
sufficient approximation. The coefficients $C_{\{ \eta\} }$ can be
obtained by solving the Bose-Hubbard Hamiltonian, or equivalently
obtained by minimizing the total energy under the normalization
constraint $ \sum {_{\{ \eta\} } \left| {C_{\{ \eta\} } }
\right|^2 } = 1$.

 The number fluctuations on a single site, the density
matrix elements, and the condensate fraction in the thermodynamic
limit can be readily calculated for this state (Appendix A). We
find:
\begin{eqnarray}
 \Delta n ^2  &=& ({M_1 }/M)(1 - {M_1 }/M),\nonumber\\
\langle \hat a_i ^\dag  \hat a_i \rangle  &=& n+{M_1 }/M,\nonumber\\
\langle \hat a_i ^\dag  \hat a_{j\ne i} \rangle  &=& {(n +
1)\Delta n^2 },\nonumber\\
f_c  &=& {\langle \hat a_i ^\dag  \hat a_{j\ne i}
\rangle}/{\langle \hat a_i ^\dag \hat a_i
\rangle}.\label{eqn:CDMX}
\end{eqnarray}
Thus, for the fixed-number state, the condensate fraction is the
ratio of the off-diagonal elements of the density matrix to the
diagonal one; the former is proportional to the number
fluctuations and the latter is just the boson density. For the
fixed-number state, as stated in Eq.(\ref{eqn:CDMX}), the
off-diagonal elements are constant, reflecting a complete
correlation within the entire system.

 If ${M_1}=0$ or ${M_1}=M$, $\Delta n ^2$ and $f_c$ both vanish and the system
is in the commensurate-filling Mott-insulator state $\left|
{\left| n \right\rangle } \right\rangle$ or $ \left| {\left| n+1
\right\rangle } \right\rangle$. If $ 0< {M_1} / M <1$, $\Delta n
^2$ and $f_c$ are both nonzero and the system is hence a
condensate with a single-particle state macroscopically occupied;
this state is the zero quasimomentum state (Appendix A). The
condensate fraction has its maximum value $(n + 1)/(\sqrt {n + 1}
+ \sqrt n )^2$ when $ M_1 /M = \sqrt {n(n + 1)} - n$ (the
difference between $n$ and the geometric mean of $n$ and $n+1$).
Compared to $100\%$ condensation of the superfluid state in the
shallow lattice limit (represented by $ [\sum_{i = 1}^M {\hat a_i
^\dag ]^N } \left| {vac} \right\rangle$), the condensate fraction
in the deep lattice regime is less than $35\%$ for all $n>0$.

\subsection{Decoupled-site mean-field ground state}\label{sec:IIB}
In comparison with the fixed-number state above, we now
investigate properties of a commonly employed mean-field ground
state. The mean-field approach introduces a scalar field to
represent the average effect of neighboring sites and leads to a
site-decoupled Hamiltonian but one which does not conserve
particle number. In particular, the Gutzwiller approximation
assumes the ground state of the system has the form $ \left| {\Psi
_{MF} } \right\rangle = \prod_{i = 1}^M {\left( {\sum_{n =
0}^\infty {f_n^{(i)} } \left| n \right\rangle _i } \right)}$,
where the ${f_n^{(i)} }$ is the amplitude of finding $n$ particles
on site $i$~\cite{Rokhsar91}. The coefficients ${f_n^{(i)} }$ can
be determined by variation to minimize the total energy $
\left\langle {\Psi _{MF} } \right|\hat H_{BH} \left| {\Psi _{MF} }
\right\rangle $. In contrast to the state of Eq. (\ref{eqn:CGS}),
$\left| {\Psi _{MF} } \right\rangle$ is neither an fixed-number
state nor a fixed-number state (i.e., its total number fluctuation
$\Delta N^2$ is nonzero; $\Delta N^2 \equiv \langle {\hat N^2 }
\rangle - \langle {\hat N} \rangle ^2$ and $ \hat N \equiv \sum_i
{\hat n_i } $). Assuming a Gutzwiller state, by substituting $\hat
a_i ^\dag  \hat a_j = \langle {\hat a_i ^\dag } \rangle \hat a_j +
\langle {\hat a_j } \rangle \hat a_i ^\dag - \langle {\hat a_i
^\dag } \rangle \langle {\hat a_j } \rangle$ into the Bose-Hubbard
Hamiltonian and diagonalizing it, one obtains the mean-field
excited states~\cite{Sheshadri93}.

In the deep lattice regime, when $ [(n - 1)U + nZJ] < \mu  < [nU -
(n + 1)ZJ] $, where $Z$ is the coordination number of the lattice,
the Gutzwiller state is $\left| {\left| n \right\rangle }
\right\rangle+{\rm O}(J/U)$ with zero condensate fraction. In this
range of the chemical potential, the system has zero
compressibility (defined as the change in the expectation value of
the total number of particles with respect to the chemical
potential or $\partial \langle {\hat N} \rangle /\partial \mu $).
This state is identified as the Mott-insulator. When $[nU - (n +
1)ZJ] < \mu < [nU + (n+1)ZJ] $, the Gutzwiller state is $ \prod
{{}_i} {\left| \psi \right\rangle _i }+ {O}(J/U)$ with $ \left|
\psi \right\rangle = f_n \left| n \right\rangle + f_{n + 1} \left|
{n + 1} \right\rangle$. Because each single-site state is a linear
combination of only two states, the system can be interpreted in
the language of spin physics (the pseudo-spin
treatment~\cite{Bruder93,Sachdev99,Altman02,Barankov07}). The
single-site state can be rewritten as
\begin{equation}
\left| \psi \right\rangle = e^{i\phi/2}\sin \frac{\theta
}{2}\left| n \right\rangle + e^{-i\phi/2}\cos \frac{\theta
}{2}\left| {n + 1} \right\rangle + {O}(J/U),
 \label{eqn:MFS}
\end{equation}
where the parameters $\theta$ and $\phi$ represent the spherical angles of the
expectation value of the spin. In the pseudo-spin model, the Mott-insulator
state is represented by a state where the pseudo-spins are either all up or all
down along the $z$ direction. The condensate state is represented by a state
where the pseudo-spins point in a $\theta$ direction determined by the chemical
potential and a $\phi$ direction determined by spontaneous symmetry breaking.
We set $\phi=0$ for convenience and find that the relation between $\theta$ and
the chemical potential is $ \cos \theta = (\mu - nU)/[(n + 1)ZJ]$. The number
fluctuations on a single site, the density matrix elements, and the condensate
fraction when $0<\theta<\pi$ are found to be:
\begin{eqnarray}
 \Delta n ^2  &=&\cos ^2 (\theta /2)\sin ^2 (\theta /2),\nonumber\\
\langle \hat a_i ^\dag  \hat a_i \rangle  &=& n+\cos ^2 (\theta /2),\nonumber\\
\langle \hat a_i ^\dag  \hat a_{j\ne i} \rangle  &=&
{(n + 1)\Delta n^2 },\nonumber\\
f_c  &=& {\langle \hat a_i ^\dag  \hat a_{j\ne i}
\rangle}/{\langle \hat a_i ^\dag \hat a_i \rangle}.
\label{eqn:MFCDMX}
\end{eqnarray}

By setting the chemical potential to the value which renders the
expectation value of the total number of particles of the
mean-field ground state equal to the total number of particles of
the fixed-number ground state, we find the relation $\cos ^2
(\theta /2) = M_1 /M$. In turn, this relationship equates the
density matrix of the two states given by Eq. (\ref{eqn:CDMX}) and
Eq. (\ref{eqn:MFCDMX}), thus showing that these two states concur
in crucial aspects of the condensate and justifying the usage of
the mean-field state for calculational purposes. In our subsequent
calculations, we therefore use whichever representation of the
condensate phase is most convenient.

\begin{figure}[t]
\centering
    \includegraphics[width=7cm]{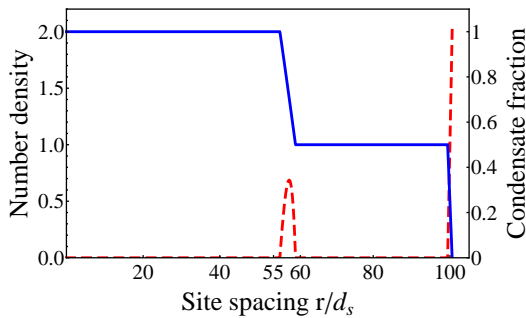}
  \caption{(Color online) Number density profile (solid line; scale on left axis
  of graph) and condensate fraction (dashed line; scale on right axis
  of graph) of an inhomogeneous lattice boson system. From inner to outer of the
  system, the phases (thickness in lattice spacing $d_s$) are the Mott-insulator (55),
  condensate (5), the Mott-insulator (38), and condensate (2).}
        \label{fig:f01}
\end{figure}

\subsection{Inhomogeneous system}
For inhomogeneous lattices, local condensate order can be studied
by replacing the chemical potential $\mu$ with $(\mu-V_i)$,
corresponding to the local density approximation. Because of the
spatial variance of $V_i$, the system is expected to display
Mott-insulator phases and condensate phases that are spatially
separated~\cite{Fisher89,Jaksch98}. In a three-dimensional (3D)
deep lattice, for example, if the potential is a radially
symmetric harmonic trap ($V_i= m\omega^2_t r_i^2/2$), the system
will have a concentric shell structure with thick Mott-insulator
shells alternating with thin condensate shells. The density
profile of this system resembles the side view of a wedding cake
(Fig.~\ref{fig:f01}).

In Fig.~\ref{fig:f01}, we plot the profile for an experimentally
feasible system containing $N=5.11\times10^6~^{87}$Rb atoms with
$a_s=5.32$~nm in a sphere whose radius is $53.2~\mu$m, or 100
sites, long. The parameters of the optical lattice are
$\lambda=1.064~\mu$m and $V_0=20E_R$ and the frequency of the
external harmonic trap is $\omega_{t}=2\pi \times 6.13$~Hz. The
system has an $n=2$ Mott-insulating sphere, an $n=1$
Mott-insulating shell, a condensate shell between the two
Mott-insulators, and a condensate shell on the outer boundary. The
thickness of the condensate interlayer is $2.66\mu$m or 5 sites
long, approximately ten times thinner than the Mott-insulating
regime. We use this setting to simulate the rf spectroscopy
experiment in Sec. III.

\section{rf spectroscopy}\label{sec:III}

One effective probe of the structure of lattice bosons is the
high-resolution microwave spectroscopy. In recent experiments, an
external rf is applied to induce $^{87}$Rb atoms to make hyperfine
transitions from $\left| F=1, m_F=-1 \right\rangle$ state to
$\left| F=2, m_F=1 \right\rangle$
state~\cite{Matthews98,Campbell06} or to $\left| F=2, m_F=0
\right\rangle$ state~\cite{Folling06}. In this section, we propose
variants of the experiment of Ref.~\cite{Campbell06} in which the
Mott-insulating shells with different occupation numbers have been
resolved by analyzing the density-dependent clock shifts of the
system. The resolution hinges on the energy differences of
interaction among two atoms in the same hyperfine state versus in
different hyperfine states.

In order to probe the condensate in the deep-lattice regime, we
propose specific values of lattice parameters that would
distinguish the spectrum of the Mott-insulator state and that of
the condensate state in two ways. The first is that the former
would have one resonant peak, but the latter would have two. The
second is that the two peaks of the latter would be blue-shifted
with respect to that of the former. The scales for the separation
between the two peaks and the shift are of a order of a few Hz,
and we anticipate that they will be within experimental reach in
the near future, providing an effective means to verify the
existence of the condensed interlayers. By considering the effect
of Goldstone modes and finite-temperature, we show that this
signal is a signature of condensate order, and would be absent if
the interlayers were a normal boson fluid of the same density.

Below, we begin with a discussion of our rf setup applied to a
homogeneous system and then use these results to derive the
spectrum of the inhomogeneous system in the local-density
approximation. We use numerical simulations where appropriate and
also invoke Fermi's golden rule to analyze the rf spectra for a
range of phase space. To summarize our findings: at
zero-temperature, the Mott-insulator and condensate states can be
clearly distinguished via their single-peak versus double-peak
structures. For the set of proposed parameters, rf transitions
into states containing Goldstone mode excitations (which are the
low-energy excitations associate with the condensate) do not
obscure the double-peak structure at zero temperature. However, at
finite temperature, thermal excitations affect the two-peak
structure. At very low temperatures ($k_B T\ll ZJ$), Goldstone
modes lead to a temperature-dependent broadening of the peaks in
the condensate signal. At higher temperatures ($k_B T\sim ZJ\ll
U$), thermal fluctuations destroy the condensate order to yield a
normal fluid; $k_BT_c$ is of order $ZJ$~\cite{Barankov07,
Gerbier07}. In this temperature regime, Goldstone modes completely
obliterate the peaks, and extra structure develops at other
characteristic frequencies corresponding to new allowed hyperfine
transitions. At temperatures $k_BT>U$, the system displays large
density fluctuations, even in the regions that are Mott-insulating
at zero temperature, and the system loses signs of its quantum
phases. Our analysis is thus limited to temperatures much less
than the interparticle interaction, $U \sim k_B\times 10$~nK
wherein, as a function of temperature, the resolution of the peaks
in the rf spectrum tracks the condensate order and its ultimate
destruction at $T_c$.

\subsection{Zero-temperature rf spectrum: Decoupled-sites}\label{sec:IIIA}

\begin{figure}[t]
\centering
    \includegraphics[width=8cm]{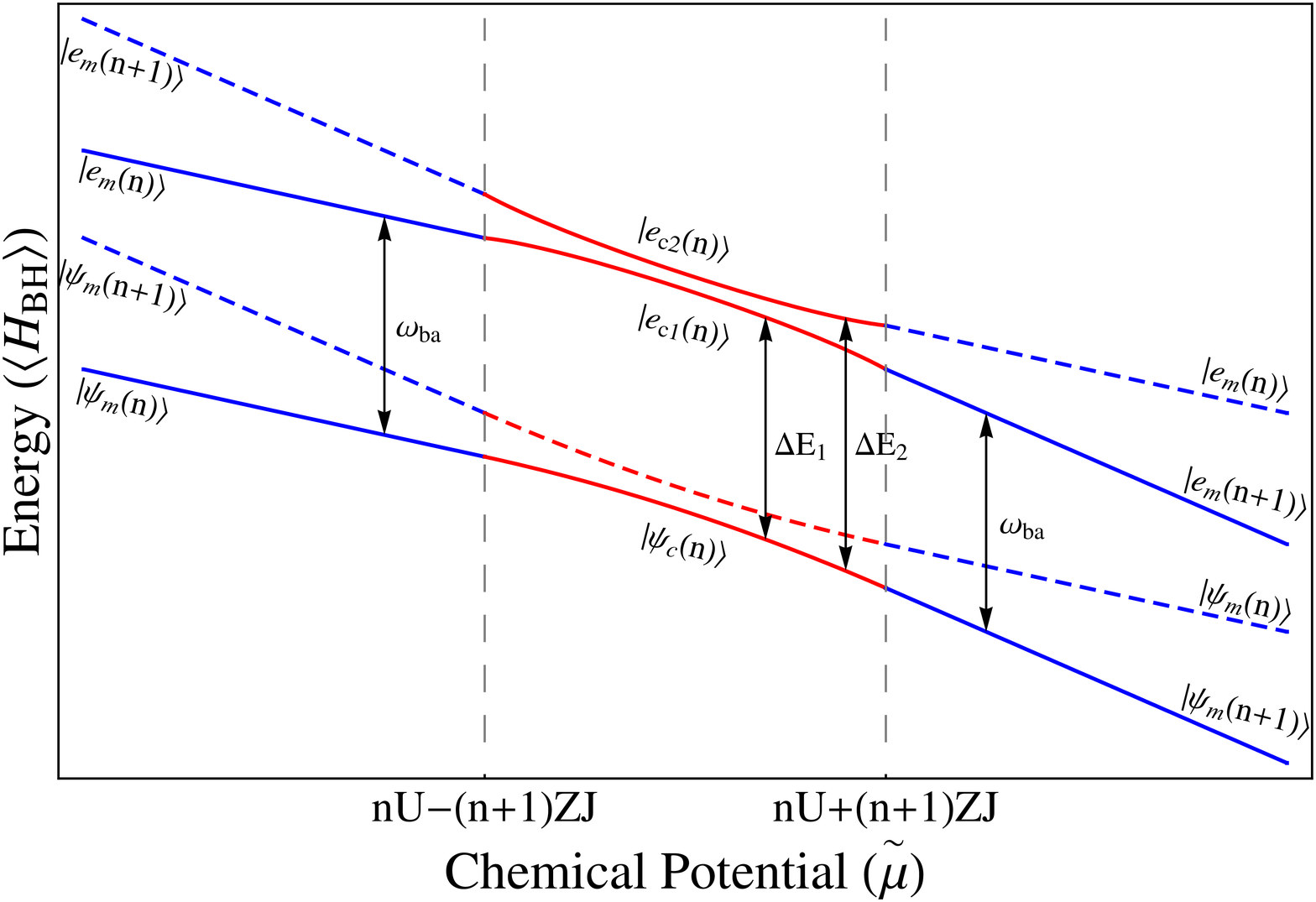}
  \caption{(Color online) Energies of the lowest four single-site
  states of the 2-species Bose-Hubbard Hamiltonian. The corresponding
  states, which are defined in Eqs. (\ref{eqn:2SGS}-\ref{eqn:CES2}),
  are marked for each curve. In the mid-region (marked red), the ground state of the
system is a condensate, while on the sides (marked blue), it is a
Mott-insulator.In each region, the ground state can make
transitions to those higher
  energy states denoted with solid lines (with corresponding energy gaps
  marked by the black arrows), but is forbidden from transitions to
  those states marked with dashed lines.
  }
    \label{fig:f02}
\end{figure}

For a uniform system of two-state ($| a \rangle$ and $| b\rangle$)
bosons, the Bose-Hubbard Hamiltonian contains two tunneling
strengths ($J_a$ and $J_b$), three interaction strengths ($U_a$,
$U_b$, and $U_{ab}$), the chemical potential, and the
single-particle energy difference between $a$ and $b$ particles
($\omega_{ba}$). In the following, we consider a particular case:
$J_b=0$, $U_b \gg U_a$, $U_{ab}=U_a$, and $\omega_{ba} \gg U_a$.
This setting has two advantages. (1) Because $\omega_{ba}>0$, the
ground state of the system has all particles in the $a$ hyperfine
state but no particles in the $b$ state. (2) Because $U_b \gg
U_{ab}=U_a$, the energy gap between having two $b$ particles on a
site and having one $b$ particle is much larger than that between
having one $b$ particle on a site and having no $b$ particles.
Therefore when the frequency of the rf field is of the order of
the gap in the latter case, we can safely limit consideration to
transitions to states with one $b$ particle.

For the situation described above, the Bose-Hubbard Hamiltonian,
including the upper hyperfine state, becomes
\begin{eqnarray}
&{}& -J\sum_{ < ij > } (\hat{a}_i^\dag \hat{a}_j + \hat{a}_j^\dag
\hat{a}_i )+\sum_i[ \frac{U}{2}(\hat{n}_{ai}+\hat{n}_{bi})
(\hat{n}_{ai}+\hat{n}_{bi}
- 1)\nonumber\\
&{}&-{\mu}(\hat{n}_{ai}+\hat{n}_{bi})
+\frac{U_b-U}{2}\hat{n}_{bi}(\hat{n}_{bi}-1)+\omega_{ba}\hat{n}_{bi}],
\label{eqn:2SBH}
\end{eqnarray}
where we drop the subscript $a$ of $J_a$ and $U_a$. In the deep
lattice regime, we use the mean-field approximation to analyze the
rf spectrum of the system. The single-site Mott-insulator state
with $\left\langle {\hat n} \right\rangle = n $ and the
single-site condensate state with $n<\left\langle {\hat n}
\right\rangle < n+1 $ can be represented  by
\begin{eqnarray}
\left| {\psi _m(n)} \right\rangle  &=& \left| n \right\rangle
\otimes \left| 0 \right\rangle,\nonumber\\
\left| {\psi _c(n) } \right\rangle  &=& \sin \frac{\theta
}{2}\left| n \right\rangle  \otimes \left| 0 \right\rangle  + \cos
\frac{\theta }{2}\left| {n + 1} \right\rangle  \otimes \left| 0
\right\rangle, \label{eqn:2SGS}
\end{eqnarray}
where $\left|{n_a}\right\rangle\otimes\left|{n_b }\right\rangle$
represents a single-site state with $n_a$ particles of hyperfine
state $a$ and $n_b$ particles of hyperfine state $b$. As in
Eq.(\ref{eqn:MES}) of Sec.~\ref{sec:IIB}, we have $\cos \theta =
(\mu  - nU)/[(n + 1)ZJ]$ determining the angle between the
pseudo-spin and the $z$ axis. Compared to Eq.(\ref{eqn:MES}), the
symmetry-breaking phase $\phi$ has been set to zero for the
decoupled-site analysis; the subsequent Goldstone mode analysis
implicitly assumes that this phase has gradual variations from
site to site.

\begin{figure}[t]
\centering
    \includegraphics[width=7cm]{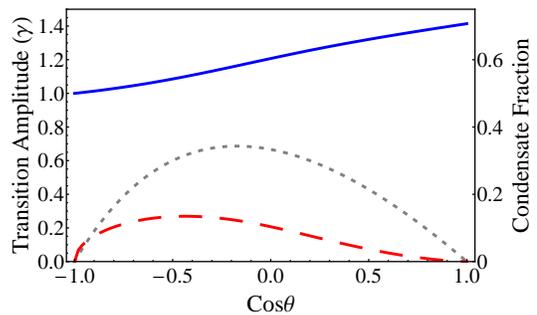}
  \caption{(Color online) Absolute value of transition amplitude $A_{1t}$ and
  $A_{2t}$ (solid and dashed lines, respectively; axis on left-hand side
  of graph) and condensate fraction (dotted line; axis on
  right-hand side of graph) as a function of the value of the cosine of the
  pseudo-spin angle (corresponding to the density of bosons).
  The system is a uniform condensate whose
  number density satisfies $1<{\left\langle n \right\rangle }<2$.}
    \label{fig:f03}
\end{figure}

The Hamiltonian describing the interaction between the bosons and
an applied rf field is~\cite{DeMarco05}
\begin{equation}
{\cal H}_{rf} = \sum_i \gamma(\hat{a}_i^\dag \hat{b}_i +
\hat{b}_i^\dag \hat{a}_i )\cos \omega t, \label{eqn:RFHam}
\end{equation}
 where $\gamma$ is proportional to
the amplitude of the rf field and $\omega$ is its frequency; this
Hamiltonian can be derived from the second-quantized form of the
interaction
$\int{d}\mathbf{r}\psi^{\dagger}\hat{\mathbf{\mu}}\cdot\mathbf{B}\psi$.

We analyze the allowed transitions in different ranges of the
chemical potential. We evaluate the transition amplitudes
$A_t=\left\langle F \right| \sum_i \gamma(\hat{a}_i^\dag \hat{b}_i
+ \hat{b}_i^\dag \hat{a}_i )\left| I \right\rangle$ between
allowed initial and final states $\left| I \right\rangle$ and
$\left| F \right\rangle$, respectively, and the energy gaps
$\Delta E$, given by the difference in the expectation values of
the Bose-Hubbard Hamiltonian of Eq. (\ref{eqn:2SBH}) in the
corresponding excited state and ground state. We plot the energy
levels in Fig.~\ref{fig:f02}, which shows how the system at zero
temperature undergoes quantum phase transitions through the $n$
Mott-insulator state [$|{\psi _m(n)} \rangle$], the condensed
state [$|{\psi _c(n)} \rangle$], and the $n+1$ Mott state [$|{\psi
_m(n+1)} \rangle$] as the chemical potential is increased. The
figure also shows possible excited states to which the ground
state can make a transition in the presence of the rf field and
shows the energy gap between them. Details of the transitions for
each range of the chemical potential in Fig.~\ref{fig:f02} are
discussed below.

When $[(n-1)U + nZJ] < \mu < [nU - (n+1)ZJ]$, the ground state is
the $n$ Mott-insulator state $[ \left| {\psi _m (n)}
\right\rangle]$. We have only one possible excited state [denoted
by $\left| e_{m}(n) \right\rangle $] to which the ground state can
make a transition. The excited state, the transition amplitude
$A_t$, and the energy gap $\Delta E$ are correspondingly
\begin{equation}
\left| e_{m}(n) \right\rangle  = \left| {n - 1} \right\rangle
\otimes \left| 1 \right\rangle,{\textrm{ }} A_t =
\gamma\sqrt{n},{\textrm{ }}\Delta E = \omega_{ba}. \label{eqn:MES}
\end{equation}
Because $\Delta E$ is independent of $n$, all the Mott-insulator
states with different $n$ have the same energy gap between the
ground state and the excited state.

When $[nU-(n+1)ZJ] < \mu <[nU+(n+1)ZJ]$, the ground state is the
condensate state $\left| {\psi _c (n)} \right\rangle$. We find two
orthogonal excited states [denoted by $\left| {e_{c1} (n)}
\right\rangle$ and $\left| {e_{c2} (n)} \right\rangle$] with
nonzero transition amplitudes. The first state, its corresponding
transition amplitude $A_{t1}$, and energy gap $\Delta E_{1}$ are
\begin{eqnarray}
\left| {e_{c1} (n)} \right\rangle  &=& \sin \frac{{\theta _1
}}{2}\left| {n - 1} \right\rangle  \otimes \left| 1 \right\rangle
+ \cos \frac{{\theta _1 }}{2}\left| n \right\rangle  \otimes
\left| 1 \right\rangle ,\nonumber\\
A_{1t}  &=& \gamma (\sqrt n \sin \frac{\theta }{2}\sin
\frac{{\theta _1 }}{2} + \sqrt {n + 1} \cos \frac{\theta }{2}\cos
\frac{{\theta
_1}}{2}),\nonumber\\
\Delta E_1 &=& \omega _{ba}  + \frac{{ZJ\sqrt {n + 1} }}{4}[\sqrt
{n
+ 1} (\sin ^2 \theta  + 2)\nonumber\\
&-& (\sqrt n \sin \theta \sin \theta _1 + 2\sqrt {n + 1} \cos
\theta \cos \theta _1 )], \label{eqn:CES1}
\end{eqnarray}
where $\left| e_{c1} \right\rangle$ is taken to be the equilibrium
state (the lowest energy state) of the Hamiltonian with the
constraint $n_b = 1$ (i.e., in the $n_b=1$ block). We calculate
the parameter $\theta_1$ by minimizing the energy of the system
with site $i$ in an excited state [$\left| {e_{c1} (n)}
\right\rangle$] and all the other sites still in the ground state
[$\left| {\psi _c (n)} \right\rangle$]. As a result, we obtain a
relation between $\theta_1$ and $\theta$, which is $ \tan \theta
_1 = \sqrt {n/(n + 1)} \tan \theta$. The second excited state,
corresponding transition amplitude $A_{t2}$, and energy gap
$\Delta E_{2}$ are
\begin{eqnarray}
\left| {e_{c2} (n)} \right\rangle  &=& \cos \frac{{\theta _1
}}{2}\left| {n - 1} \right\rangle  \otimes \left| 1 \right\rangle
- \sin \frac{{\theta _1 }}{2}\left| n \right\rangle  \otimes
\left| 1 \right\rangle ,\nonumber\\
A_{2t}  &=& \gamma (\sqrt n \sin \frac{\theta }{2}\cos
\frac{{\theta _1 }}{2} - \sqrt {n + 1} \cos \frac{\theta }{2}\sin
\frac{{\theta
_1}}{2}),\nonumber\\
\Delta E_2 &=& \omega_{ba}  + \frac{{ZJ\sqrt {n + 1} }}{4}[\sqrt
{n
+ 1} (\sin ^2 \theta  + 2)\nonumber\\
&+& (\sqrt n \sin \theta \sin \theta _1 + 2\sqrt {n + 1} \cos
\theta \cos \theta _1 )]. \label{eqn:CES2}
\end{eqnarray}

\begin{figure}[t]
\centering
    \includegraphics[width=8cm]{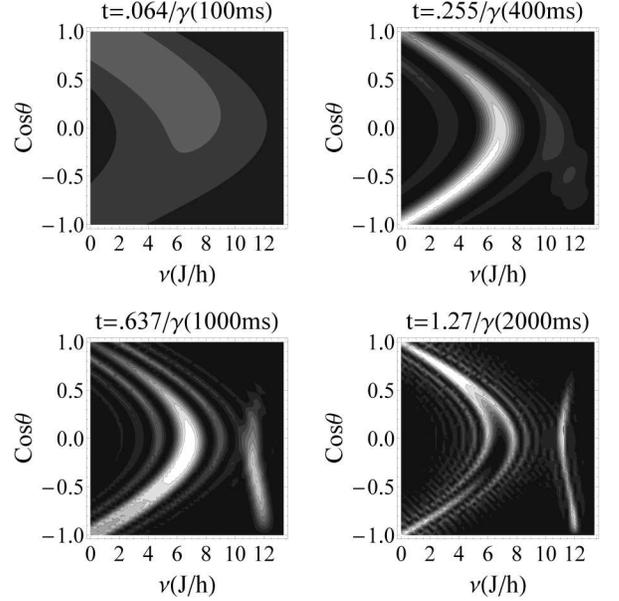}
  \caption{Fraction of bosons transferred to the higher hyperfine state for a uniform condensate with
  $1<{\left\langle n \right\rangle }<2$ at $t=100$, 400, 1000, and 2000 ms.
  The horizontal axis is the detuning of the rf field with respect
  to the resonant frequency of the Mott-insulating state, while
  the vertical axis is $\cos\theta$. The color scale is bounded by 0 (black) and 1 (white) with contours separated by 0.1.}
      \label{fig:f04}
\end{figure}

When the chemical potential in the condensed phase reaches its
upper or lower bound ($\theta=0$ or $\theta=\pi$), Eq.
(\ref{eqn:CES1}) becomes Eq. (\ref{eqn:MES}) with $n$ or $n+1$
Mott-insulator states correspondingly, but the transition
amplitude of Eq. (\ref{eqn:CES2}) vanishes~(Fig.~\ref{fig:f03}).
However, the $n=0$ condensate state is an exception since it can
only make a transition to one excited state $\left| {0}
\right\rangle \otimes \left| 1 \right\rangle$ with a transition
amplitude $\gamma\cos{\frac{\theta}{2}}$ and an energy gap
$(\omega_{ba}+JZ\sin^4\frac{\theta}{2})$.

\vspace{0.3cm} {\it Thus, while the Mott state can only make a transition to
one excited state, the condensate state can make transitions to two excited
states.}\vspace{0.3cm}

\begin{figure}[t]
\centering
    \includegraphics[width=6.5cm]{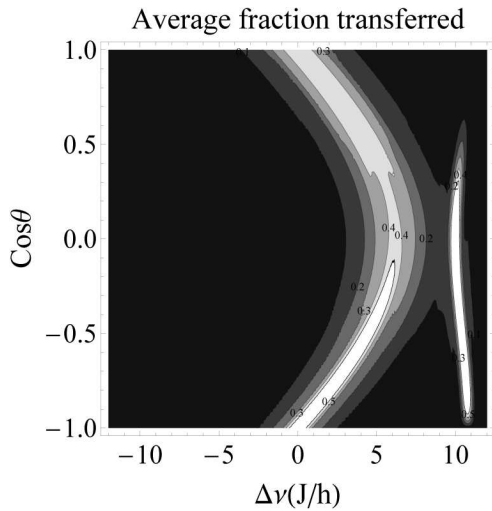}
  \caption{Average fraction of bosons transferred between $t=1000$ and $2000$~ms.
  Contour interval is $0.1$. The first peak appears in the range of
  $0$ to $7J$ ($6.29$~Hz), and the second peak appears at about $11.5J$
   ($10.3$~Hz). When $\cos\theta=\pm 1$, we recover the Mott signature.}
    \label{fig:f05}
\end{figure}

\begin{figure}[t]
\centering
    \includegraphics[width=6.5cm]{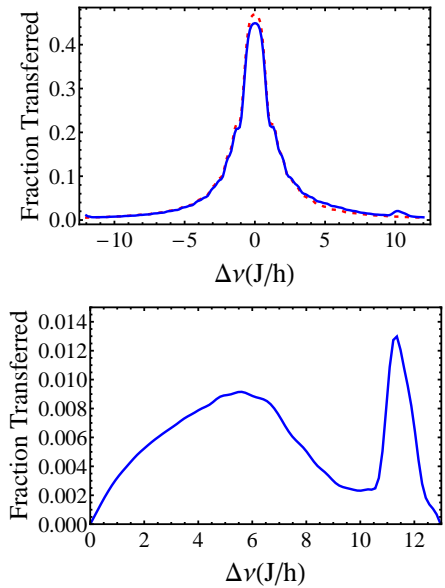}
  \caption{(Color online) Average fraction of bosons transferred
  between $t=1000$ ms and $2000$ ms for the inhomogeneous system defined
  in Fig.~\ref{fig:f01}. Up: the system with both Mott-insulator and condensate
  regions (solid) compared with that having only Mott-insulator regions (dotted). Down:
  spectrum of the condensate shells after eliminating the contribution from the Mott
 -insulator shells. The results show that two peaks with positive shifts are
  signatures of condensate order in a deep optical lattice.
  }
    \label{fig:f06}
\end{figure}

As a consequence of these allowed transitions, the Mott-insulator
state ought to have a single peak in the associated rf spectrum
while the condensate ought to have two. This can be seen to first
order (in number of particles transferred) by employing Fermi's
golden rule (FGR), where the transition rate is given by
~\cite{Pethick01}
\begin{equation}
I(\omega ) = \frac{2\pi}{\hbar}\sum_{F,I} {(\rho _I  - \rho
_F )\left| {A_t } \right|^2 } \delta \left( {\omega  - E_F + E_I }
\right), \label{eqn:FGR}
\end{equation}
where $\left| I \right\rangle$ ($\left| F \right\rangle$) is the
initial (final) state with energy $E_I$ ($E_F$) and probability
for occurrence $\rho_I$ ($\rho_F$). Hence, FGR would predict a
single delta function peak for the Mott-insulator state rf
spectrum and two delta function peaks for the condensate.

We now confirm that the presence of the condensate does yield a
two-peak rf spectrum by numerically time evolving the equations of
motion in the presence of the rf field. Taking into account both
experimental feasibility and optimal conditions for observing the
rf signatures of the condensate, we consider the setting of
Fig.~\ref{fig:f01}. This setting corresponds to the values
$U=2\pi\times 306$ Hz, $J=2\pi \times 0.898$ Hz (then
$ZJ/U=1.76\times 10^{-2}$ is small compared to $1$), and
$\mu=1.5U$. The intensity of the rf field is set such that
$\gamma=0.118ZJ=2\pi\times 0.637$ Hz, which is of the same order
as that in the experiment of Ref.~\cite{Campbell06}. If the rf
field is turned on at $t=0$, within the rotating-wave
approximation we find that the fraction of transferred particles
at the frequency of the second peak of the condensate's spectrum
begins to rise at $t=0.225/\gamma$ ($400$ ms), and that at any
given time, the spectrum shows oscillatory behavior with respect
to the detuning (Fig.~\ref{fig:f04}). In order to see a distinct
second peak, we calculate the average fraction transferred between
$t=0.637/\gamma$ ($1000$ ms) and $t=1.27/\gamma$ ($2000$ ms).
Figure~\ref{fig:f05} shows the average fraction transferred for a
uniform condensate. We can see two peaks whose positions agree
with the energy gap in Eqs. (\ref{eqn:CES1}) and (\ref{eqn:CES2}).
Figure ~\ref{fig:f06} shows the average fraction transferred for
the entire inhomogeneous system. Because the fraction transferred
in the Mott-insulator is an even function of the detuning, we
subtract the negative detuning part from the corresponding
positive detuning part to eliminate the contribution of the
Mott-insulator. The result shows that there are two peaks of
magnitude 1\% ($5\times 10^5$ atoms transferred), and their
positions agree with our theoretical analysis.

  All calculations above rely on a mean-field
approximation. As a minor check that number conservation and
entanglement do not alter the proposed signatures, we performed
the numerical toy simulation os time-evolving the entangled,
fixed-number condensate state of six bosons in four sites in the
presence of an rf field (Appendix C). Even such small system sizes
has been shown to exhibit salient features of the
superfluid-Mott-insulator phase transition in macroscopic
systems~\cite{Luhmann08}. The results show the same signatures as
those of the mean-field approach: two peaks with positive shifts
of order $J$, thus providing some confidence that our result is
not an artifact of the mean-field approximation.

To ensure that no spurious effects emerge from truncating the Hilbert space, we
have analyzed the effect of the leading subdominant states $\left| {n - 1}
\right\rangle \otimes \left| 0 \right\rangle$ and $\left| {n +2} \right\rangle
\otimes \left| 0 \right\rangle$. Within the resultant four-state truncated
space, we obtain four possible final states. Two of them have the same energy
shifts (of order $J$) and transition amplitudes (of order $\gamma$) as those
obtained in the two-state truncated space, plus small corrections of order
$J/U$; these corrections contribute a fraction $(J/U)^2$ to the number
transferred. The other two final states have energy shifts of order $U$ ($\gg
J$) and transition amplitudes of order $\gamma J/U$ ($\ll \gamma$), which also
contributes a fraction $(J/U)^2$ to the number transferred. The minimal change
coming from taking into account these additional states justifies our usage of
the two-state truncation in the deep lattice regime.

\subsection{Goldstone modes and finite-temperature effects}\label{sec:IIIB}
At finite temperature, the thermally-occupied excited states of
the system will affect the rf spectrum. When the temperature is
much smaller than the interaction energy ($k_B T\ll U$), the
Mott-insulator state has negligible thermal effects because it has
an excitation gap of order
$U$~\cite{DeMarco05,Pupillo06b,Konabe06}. However, we should be
concerned that the two-peak signature of the condensate may be
destroyed at low temperatures by Goldstone modes associated with
the continuous symmetry-breaking in the phase $\phi$ of the
condensate ground-state wave function. Each Goldstone mode or
boson corresponds to a long-wavelength distortion of the density
and phase between neighboring sites and the modes form the gapless
low-energy excitation spectrum for the condensate. The structure
of these modes has been derived from the pseudo-spin formulation
and analyzed for homogeneous and inhomogeneous systems in Ref.
\cite{Barankov07}. In order to be self-contained, we present the
Goldstone-mode description here in Appendix B for the homogeneous
generalized case of the two-species system of interest in this
paper.

At zero temperature, the initial state consists of the ground
state of the system purely comprised of $a$ particles, and thus no
Goldstone modes are excited. Under an rf field however, a small
number of particles change their internal state to $b$. If this
change is not uniform in space, it will be accompanied by
Goldstone modes. Assuming that $N_G$ is the total number of
Goldstone bosons excited by the rf field, we can estimate the
final-state energy of a single site, which differs from that of
the $N_G=0$ case, as $(N_G/M)ZJ$, where $M$ is the total number of
sites and $ZJ$ is the energy scale of the Goldstone-mode spectrum.
If the rf field is weak enough that the average number of excited
Goldstone bosons per site ($N_G/M$) is much smaller than one, the
change in the single-site energy gap is much smaller than $ZJ$,
which is about the distance between the two peaks obtained in the
Sec. II. In typical experimental settings, this indeed is the case
given that about $10\%$ of the particles make transitions to the
$b$ state. Therefore, exciting Goldstone modes at zero temperature
ought not obscure the two-peak signature; in Appendix B, we
explicitly show that these modes provide a small background which
still leaves the two-peak signature robust.

 At finite temperature, in three dimensions, the condensate density is reduced by the
existence of thermally excited Goldstone modes. As argued in
Appendix B, for low temperatures $kT\ll ZJ$, we expect a slight
reduction in the height of the two peaks and a slightly larger
contribution to the background. Nevertheless, for
three-dimensional systems, we expect the two-peak condensate
signature to persist at these temperatures.

Close to the critical temperature $kT_c\approx ZJ$ (see Ref.
~\cite{Barankov07, Gerbier07}), we expect the average thermal
expectation value $\sum{_j}\langle {a_j }\rangle/N$ of the order
parameter to vanish, corresponding to destruction of long-range
order in the system. Within the decoupled-site mean-field
approximation, $\langle {a_i }\rangle$ fluctuates in magnitude and
phase from site to site. The kinetic part of the mean-field energy
in the Bose-Hubbard Hamiltonian can thus be thought of as having a
spread of order $(n+1)ZJ$. Given that the two low-temperature
peaks in the rf spectrum are also separated by about $(n+1)ZJ$, we
predict that the two peaks merge into a continuum around the
critical temperature. In two-dimensions, even the smallest
temperature suffices for the Goldstone modes to destroy true
long-range order and we believe that this will be reflected in the
smearing out of the two-peak structure for low-dimensional
systems.

Finally, for the inhomogeneous shell situation, the Goldstone
modes become quantized due to confinement~\cite{Barankov07}.
Reference \cite{Barankov07} provides a discussion of the mode
structure and dimensional crossover as a function of temperature.
The essential issue is whether the quantized modes along each
direction which are accessible at temperature $T$ are numerous
enough to form an effective continuum. Modifying the arguments
above to include these inhomogeneous effects, we deduce that the
two peak structure would remain at low temperatures for a
condensate shell with a thickness of several lattice sites.

In summary, our arguments indicate that the presence of condensate
order can be detected in a two-peak rf spectrum for appropriate rf
parameter settings in contrast to the one-peak structure of the
Mott-insulator phase. Our results also suggest that when the
condensate becomes a normal fluid at $T=T_c$, the two-peak
structure is washed out. Our arguments can be made more rigorous
by way of numerical simulations such as those of Sec. IIIA that
would include the Goldstone modes and finite temperature effects;
such a treatment is beyond the scope of this paper.

\section{Matter wave interference}\label{sec:IV}

 Experimental observation of matter-wave interference peaks in absorption images
have provided striking evidence for superfluid states in shallow
optical lattices~\cite{Greiner02}. In addition, the concentric
Mott-insulating-shell system has been observed to preserve finite
visibility in the interference
pattern~\cite{Gerbier05a,Gerbier05b}. Characteristics of the
interference patterns have been analytically and numerically
studied under several specific conditions for normal fluid,
superfluid and Mott-insulating phases in the lattice boson
system~\cite{Roth03,Diener07,Lin08,Toth08}. Here, we consider
bosons in the deep lattice regime and ask what signatures of
matter-wave interference can distinguish the presence of a
condensate shell between two Mott-insulating shells. We work at
zero temperature, which should be valid for temperatures $kT\ll
J$, where one would expect the contrast between these states to be
strongest. We begin by considering a homogeneous system, where the
many-body wave functions of the Mott-insulating and deep-lattice
condensate phases can be exactly time-evolved to obtain the
density profile measured by absorption imaging. For simplicity, we
ignore the effect of interaction during time of
flight~\cite{Roth03}, which is expected to quantitatively
influence the intensity and width of peaks but preserve the
qualitative signatures of the interference pattern~\cite{Lin08}.
By calculating the time-evolution of a system of bosons on a ring
lattice, we show that condensate states will have sharp
interference maxima, in contrast to Mott-insulator states, after
free expansion of the system. To illustrate how these features
would be displayed in an inhomogeneous system, we also the time
evolution of a system of concentric ring lattices that have a
Mott-condensate-Mott structure and point to signatures of the
condensate in the time-evolved profiles.

\subsection{Density profile upon expansion for homogeneous systems}\label{sec:IVA}

For the Mott-insulator state $\left| {\left| n \right\rangle }
\right\rangle$, the many-body wave function is a product of $N$
single-particle wave functions:
\begin{eqnarray}
\Psi ({\bf{r}}_1 , \ldots ,{\bf{r}}_N ) &\equiv& \left\langle
{{{\bf{r}}_1 , \ldots ,{\bf{r}}_N }}
 \mathrel{\left | {\vphantom {{{\bf{r}}_1 , \ldots ,{\bf{r}}_N } {\left| n \right\rangle }}}
 \right. \kern-\nulldelimiterspace}
 {{\left| n \right\rangle }} \right\rangle \nonumber\\
 &=& A\sum_{\textrm{sym}\{ {\bf{r}}\} } {\prod_{i = 1}^N {\psi ({\bf{r}}_i  - {\bf{s}}^{(i)} )}
 }\nonumber\\
 &=& A \sum_{\{ {\bf{s}}\} }
{\prod_{i = 1}^N {\psi ({\bf{r}}_i  - {\bf{s}}^{(i)} )}
 },
\end{eqnarray}
where ${\psi ({\bf{r}}_i  - {\bf{s}}^{(i)} )}$ is the single
particle wave function of particle $i$ localized on a site at
position ${\bf{s}}^{(i)}$, and $A$ is a normalization constant.
To satisfy Bose-Einstein statistics, we need to symmetrize all $N$
degrees of freedom; this symmetrization (denoted by $\sum
{_{\textrm{sym}\{ {\bf{r}}\} } } $) is equivalent to the sum over
all configurations of $\{ {\bf{s}}^{(i)}\}$ (denoted by $\sum
{_{\{{\bf{s}}\}} } $), which are permutations of $ \{ {\underbrace
{{\bf{s}}_1 , \ldots ,{\bf{s}}_1 }_n,\underbrace {{\bf{s}}_2 ,
\ldots ,{\bf{s}}_2 }_n, \ldots \underbrace {{\bf{s}}_M , \ldots
,{\bf{s}}_M }_n} \}$, where ${{\bf{s}}_j } $ is the position of
site $j$.

Near the center of a lattice site, the lattice potential can be
approximated as harmonic with frequency
$\omega=\sqrt{4{V_0}{E_R}}/\hbar$, where $V_0$ is the depth of the
optical lattice and $E_R$ is the recoil energy. The
single-particle wave function can thus be approximated as a
harmonic oscillator ground-state wave function well-localized on a
site. If we ignore interactions (valid at low density), after
turning off the lattice potential and the trap potential all
single-particle wave functions are time-evolved by the free
particle Hamiltonian ${\bf{\hat p}}^2/2m $. The wave function at
later times is then given by
\begin{eqnarray}
\psi _t ({\bf{r}}_i  - {\bf{s}}^{(i)} ) =  \left( {\frac{{2l^2
}}{\pi }} \right)^{d/4} \frac{{\exp [ - ({\bf{r}}_i -
{\bf{s}}^{(i)} )^2 /(l^2  + 2i\hbar t/m)]}}{{(l^2 + 2i\hbar
t/m)^{d/2} }},\nonumber\\
\end{eqnarray}
where $d$ is the dimensionality of the system, and $l$ is the
characteristic length of the single-particle wave
function~\cite{Griffiths05,Toth08}. The density profile as a
function of position and time is obtained by calculating the
diagonal elements of the single-particle density
matrix~\cite{Leggett06};
\begin{eqnarray}
\rho (r,t) &\equiv& N \int {\Psi _t^* ({\bf{r}},{\bf{r}}_2 ,
\ldots ,{\bf{r}}_N )\Psi _t ({\bf{r}},{\bf{r}}_2 , \ldots
,{\bf{r}}_N
)d{\bf{r}}_2  \ldots d{\bf{r}}_N }\nonumber\\
&=& N\left| A \right|^2 \sum_{\{ {\bf{s}}\} }
{\sum_{\{ {\bf{s'}}\} } {\psi _t^* ({\bf{r}} -
{\bf{s}}^{(1)} )\psi _t ({\bf{r}} - {\bf{s'}}^{(1)} )} }
\nonumber\\
&{ }& \times \exp [ - \sum_{i = 2}^N {({\bf{s}}^{(i)}  -
{\bf{s'}}^{(i)} )^2 /(2l^2 )} ]. \label{eqn:MDP}
\end{eqnarray}

For the condensate state $\left| \Psi \right\rangle$ of Eq.
(\ref{eqn:CGS}), the density profile is given by
\begin{eqnarray}
 \rho _c (r,t) &=& N\left| {A} \right|^2 \sum_{\{ \eta\} }
{\sum_{\{ \eta'\} } {C_{\{ \eta'\} }^* C_{\{ \eta\} } } }\nonumber\\
 &\times& \{ {\sum_{\{ {\bf{s}};\{ \eta\} \} } {\sum_{\{ {\bf{s'}};\{ \eta'\} \} }
  {\psi _t^* ({\bf{r}} - {\bf{s}}^{(1)} )\psi _t ({\bf{r}} - {\bf{s'}}^{(1)} )} } }
 \nonumber\\
  &\times& \exp [ - \sum_{i = 2}^N {({\bf{s}}^{(i)} -
{\bf{s'}}^{(i)} )^2 /(2l^2 )} ]\},
 \label{eqn:CDP}
\end{eqnarray}
where $\sum {_{\{ \eta\}} }$ denotes the same sum in Eq.
(\ref{eqn:CGS}), and $\sum {_{\{\bf{s};\{ \eta\}\}} }$ denotes the
sum over all $\{ {\bf{s}}^{(i)}\}$ with a specific ${\{ \eta\}}$;
$\{ {\bf{s}}^{(i)}\}$ are permutations of the set including $n+1$
each ${\bf{s}}_{\eta_i}$ of the sites $\eta_1,\ldots ,\eta_{M_1}$
and $n$ each ${\bf{s}}_{j}$ of the other sites. If $M_1 = 0$ or
$M_1=M$, ${\{ \eta\}}$ has only one configuration, and hence Eq.
(\ref{eqn:CDP}) becomes Eq. (\ref{eqn:MDP}).

\subsection{Example: Expansion of bosons in a ring lattice}\label{sec:IVB}

As a concrete case that shows how sharp interference peaks emerge upon release
and expansion of the condensate, we consider the illustrative example of bosons
in a one-dimensional lattice of sites located on a ring of radius $s$ in a two
dimensional plane. Because of the symmetry of the ring, constructive
interference is expected at its center. We therefore calculate the center
density contributed by one particle, defined as $ \tilde \rho (t) \equiv \rho
(0,t)/N$. The result is
\begin{equation}
\tilde \rho (t) = F\left( {\frac{{2l^2 }}{\pi }} \right)^{d/2}
\frac{{\exp [ - 2s^2 l^2 /(l^4  + 4\hbar ^2 t^2 /m^2 )]}}{{(l^4  +
4\hbar ^2 t^2 /m^2 )^{d/2} }}, \label{eqn:RCDP}
\end{equation}
where $d=2$ for this system and $F$ is equal to
\begin{eqnarray}
F_{M} = \left| A \right|^2 \sum_{\{ {\bf{s}}\} } {\sum_{\{
{\bf{s'}}\} } {\exp [ - \sum_{i = 2}^N {({\bf{s}}^{(i)}  -
{\bf{s'}}^{(i)} )^2 /(2l^2 )} } } ]\nonumber
\end{eqnarray}
for the Mott-insulator state and
\begin{eqnarray}
F_{c} &=&  \left| A \right|^2 \sum_{\{ \eta\} } {\sum_{\{ \eta'\}
}
{C_{\{ \eta'\} }^* C_{\{ \eta\} }  } }\nonumber\\
& & \times \{ \sum_{\{ {\bf{s}};\{ \eta\} \} } {\sum_{\{
{\bf{s'}};\{ \eta'\} \} } {\exp [ - \sum_{i = 2}^N
{({\bf{s}}^{(i)} - {\bf{s'}}^{(i)} )^2 /(2l^2 )} ]} } \}\nonumber
\end{eqnarray}
for the condensate. If $s \gg l$, $\tilde \rho (t)$ has a maximum
at
\begin{equation}
t_m = mls/\sqrt{d}\hbar \label{eqn:PT}
\end{equation}
and a half width $\Delta{t}=2^{1/d} mls/\hbar $. While $t_m$
indicates the time it takes matter waves to disperse from the ring
to the center, $\Delta{t}$ measures the duration of constructive
interference processes. Equation (\ref{eqn:PT}) is of the same
form as
 a continuum superfluid confined to a shell geometry with the characteristic
length $l$ given by the thickness of the shell~\cite{Lannert07}.
The terms with $\{ \eta'\}=\{ \eta\}$ in the sum for $F$ are the
same for the Mott-insulator and condensate states. However, $F$
for the condensate state has many additional terms with $\{
\eta'\}\neq\{ \eta\}$ that contribute to the center density which
are not present in $F_{M}$. Therefore, we see that the center
density of the condensate state has a much sharper peak than that
of the Mott-insulator state. At finite temperature, the
interference pattern is the thermal average over the densities of
all possible pure states. The coefficients $ C_{\{ \eta\} }$ of
higher energy states are not necessarily real and positive and
thus would decrease the value of $F$ given that not all terms give
positive contributions. Therefore, we expect that the interference
pattern of a condensate becomes more blurred with increasing
temperature.

Although here we have only analytically solved for the density at
the center of the ring, constructive interference patterns will
occur throughout space after the system is released from its ring
trap. The differences in these spatial patterns between initial
Mott-insulator and initial condensed states gives further evidence
for condensate order. As a toy example that exhibits these
patterns, we numerically simulated the expansion processes of a
four-site Mott-insulator and a four-site condensate in the ring
geometry (Appendix C). Although the initial density profiles of
the two cases look the same, their interference patterns behave
quite differently during the expansion process.

\subsection{Inhomogeneous systems}\label{sec:IVC}

\begin{figure}[t]
\centering
    \includegraphics[width=7cm]{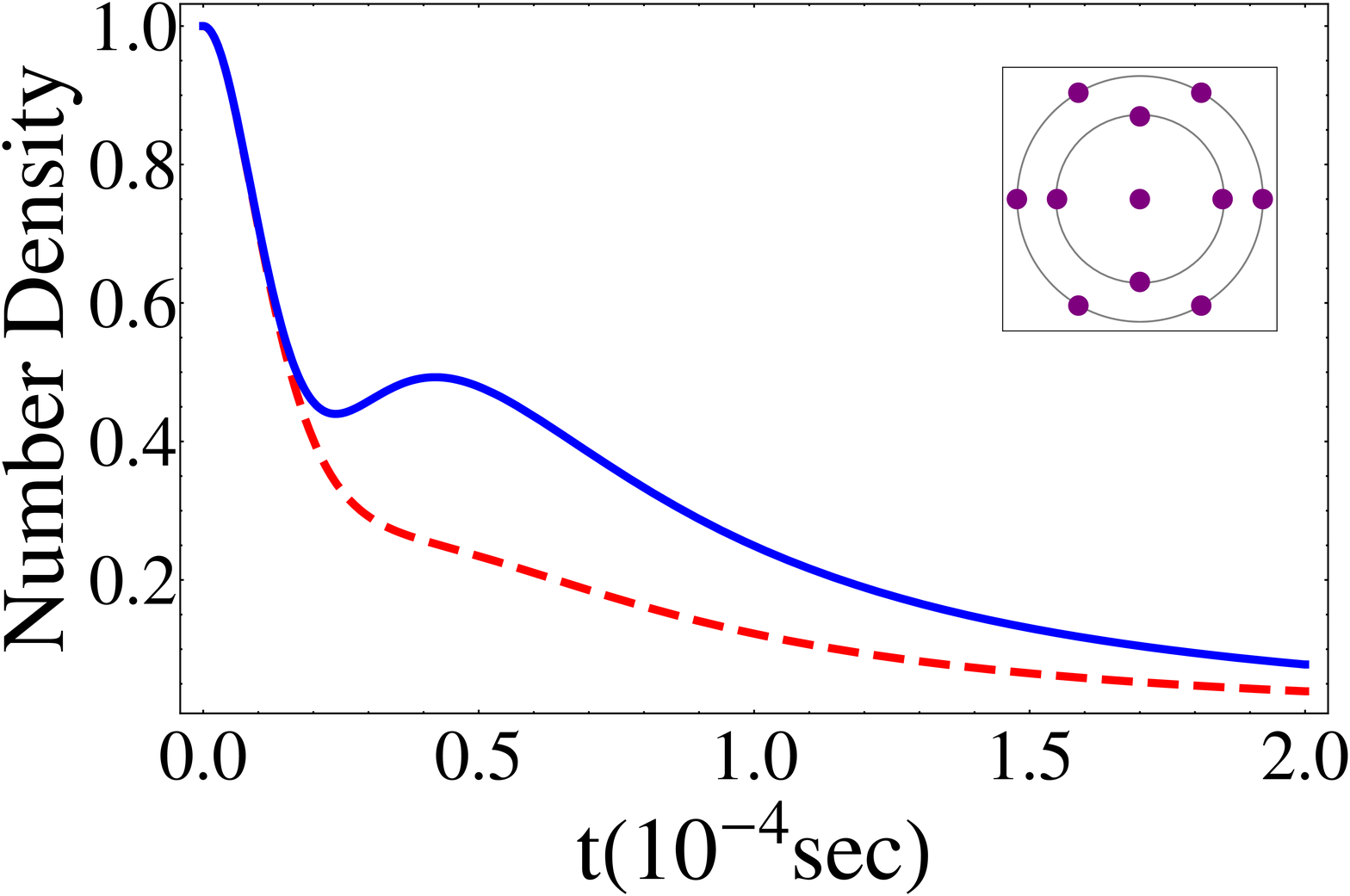}
  \caption{(Color online) Density at the center point, as a function of time, during expansion of the system. The inset shows
  the geometry and the site positions of the model. The solid curve
  corresponds to the system that has a center site and an outer
  ring in Mott-insulating states and the inner ring in the condensate
  state. The dashed curve corresponds to the control system with the inner ring in
  the Mott-insulating state. The vertical axis is the relative density with
  respect to the initial value. The peak and the relatively large value at long times
  are evidence of the condensate part of the system.}
    \label{fig:f07}
\end{figure}

The above discussions show how homogeneous systems of condensate
and Mott-insulator display different patterns upon release and
expansion and that interference is a signature of condensate order
in the system. To illustrate how these signatures can be used to
probe condensate order realistic experimental situations, we
consider a simplified inhomogeneous system of  $^{87}$Rb atoms
displaying the nested, or ``wedding-cake," structure of bosons in
optical lattices in the presence of a harmonic trap. Building on
the results of the previous section for bosons on a ring lattice,
we model a two-dimensional system made up of a center point and
two concentric rings, where the central site is in the $n=2$
Mott-insulating state, the four sites in the inner ring are in the
${\left\langle n \right\rangle=1.5}$ condensate state,  and six
sites in the outer ring are in the $n=1$ Mott-insulating state.
Site spacings of the two rings are both $d_s=0.532~\mu$m. For the
lattice depth $V_0=20E_R$, the characteristic length of the
initial single-particle wave function is $l=0.283d_s$. If the
lattice potential is turned off at $t=0$, the density at the
center point as a function of time shows a peak at
$t=4.2\times10^{-5}$~s (Fig.~\ref{fig:f07}). Compared to a control
system where the inner ring has Mott-insulator states of the same
density ($2$ sites in $n=2$ and $2$ sites in $n=1$), the peak and
the larger long-time residual density in the center indicate large
constructive interference due to the condensate part of the
system. Considering expansion of a 3D inhomogeneous system of
Fig.~\ref{fig:f01}, we expect to observe a density peak in the
center of the sphere at $t=3.67$~ms, which is about $1/4$ the
duration of the time of flight in the experiment of
Ref.~\cite{Greiner02}. We expect that such a time-evolved density
profile should also be able to discern the presence of condensate
interlayers for larger systems with realistic numbers of bosons.

In summary, we obtain interference signatures of a thin condensate
inter-layer between two Mott-insulator layers. For a toy system
exhibiting concentric co-existent phases, we show that the
condensed phase contributes a sharp peak to the time-evolved
central particle density. As performed for single phases,
simulations of realistic inhomogeneous large-sized
three-dimensional systems within our truncated wave function basis
and extraction of measurable quantities, such as the visibility,
are in order; we believe that our results here are indicative of
the signatures of the condensate layer that would be obtained in
actual systems.

\section{Summary}\label{sec:V}
Systems of atoms in optical lattices provide a seemingly-ideal
system for exploring the quantum phases of bosons. Generically,
experimentally-studied systems are not of uniform density, due to
the confining trap. This leads to the coexistence of
Mott-insulating and condensate phases within a single system at
fixed lattice depth and complicates the study of these phases.
Here, we have considered bosons in deep optical lattices, where
condensate phases exist between Mott-insulator phases, but are
close to their quantum transition to the Mott-insulator phase. We
have justified the use of a mean-field approximation by
constructing the entangled, fixed-number, many-body wave function
and showing that many key aspects of this phase (in particular,
the condensate fraction and the on-site number fluctuations) are
identical in the two descriptions. Toward developing experimental
tests of the condensate interlayers in inhomogeneous systems, we
have calculated the condensate signal using two common probes of
cold-atom systems: rf spectroscopy and release-and-expansion. We
have shown specific features of the signal in these probes that
distinguish the condensate regions from the Mott phases. We have
shown that the rf signal specifically probes condensate order in
that these characteristic features disappear when the condensate
is replaced by a normal fluid as a result of thermal fluctuations.
We hope that our work will enable closer exploration of the
structure of coexisting Mott-insulator-condensate phases in
realistic experimental situations.

\section{Acknowledgements}
We are grateful to B. DeMarco, K. R. A. Hazzard, A. J. Leggett, E.
J. Mueller, and M.-H. Yung for involved discussions and G. Baym
for his perceptive remarks. We would like to thank R. Barankov for
his input during the initial stages of this work and for his
insights into the Goldstone-mode structure in the case of a single
species of bosons. C.L. and S.V. would like to acknowledge the
gracious hospitality of the Aspen Center for Physics where some
key aspects of this research were sorted out. This work was
supported by the NSF under Grants No. DMR-0644022-CAR (S.V.) and
DMR-0605871 (C.L.).

\appendix
\section{Density matrix of the fixed-number condensate state}\label{sec:APA}

In this appendix, the elements of the density matrix of Eq.
(\ref{eqn:CGS}) are calculated. For diagonal terms,
\begin{eqnarray}
\langle \hat a_i ^\dag  \hat a_i \rangle  &=& \sum\limits_{\{
\eta\} ,\{ \eta'\} } {\frac{{C_{\{ \eta'\} }^* C_{\{ \eta\} }
}}{{(n + 1)^{M_1 } }}\left\langle {\left\langle n \right|}
\right|(\prod\limits_{j = 1}^{M_1 } {\hat a_{\eta'_j } } )\hat a_i
^\dag  \hat a_i (\prod\limits_{j = 1}^{M_1 } {\hat a_{\eta_j }
^\dag } )\left|
{\left| n \right\rangle } \right\rangle }\nonumber\\
&=& \sum\limits_{i \notin \{ \eta\} } {\left| {C_{\{ \eta\} } }
\right|^2 \langle {\hat a_i ^\dag  \hat a_i } \rangle _i } +
\sum\limits_{i \in \{ \eta\} } {\frac{{\left| {C_{\{ \eta\} } }
\right|^2 }}{{n + 1}}\langle {\hat a_i \hat a_i ^\dag  \hat a_i
\hat a_i ^\dag  } \rangle _i }
\nonumber\\
 &=& n\sum\limits_{i \notin \{ \eta\} } {\left| {C_{\{ \eta\}
} } \right|^2 }  + (n + 1)\sum\limits_{i \in \{ \eta\} } {\left|
{C_{\{ \eta\} } } \right|^2 }
\nonumber\\
&=& n + \sum\limits_{i \in \{ \eta\} } {\left| {C_{\{ \eta\} } }
\right|^2 },\nonumber
\end{eqnarray}
where $ \langle \hat a_i ^\dag  \hat a_i \rangle _i  \equiv
{}_i\left\langle n \right|\hat a_i ^\dag  \hat a_i \left| n
\right\rangle _i$. From symmetry, $\sum_{i \in \{ \eta\} } {\left|
{C_{\{ \eta\} } } \right|^2 }$ is independent of $i$ but depends
only on the number of terms in the sum. Considering the
normalization condition, we find
\begin{eqnarray}
\sum\limits_{i \in \{ \eta\} } {\left| {C_{\{ \eta\} } } \right|^2
} &=& \frac{{{\textrm{\#  of terms in the sum }}\sum {{}_{i \in \{
\eta\} }} }}{{{\textrm{\#  of terms in the sum }}\sum {{}_{\{
\eta\} }}
}}\nonumber\\
&=&
  \frac{{\left( {\begin{array}{*{20}c}
   {M - 1}  \\
   {M_1  - 1}  \\
\end{array}} \right)}}{{\left( {\begin{array}{*{20}c}
   M  \\
   {M_1 }  \\
\end{array}} \right)}} = \frac{{M_1 }}{M}\nonumber\\
 \Rightarrow \langle \hat a_i ^\dag  \hat a_i \rangle  &=& n + \frac{{M_1
 }}{M}.
\end{eqnarray}

To evaluate the number fluctuation of a single site, we need to calculate
$\langle \hat a_i ^\dag  \hat a_i \hat a_i ^\dag  \hat a_i \rangle$. Similarly,
\begin{eqnarray}
\langle \hat a_i ^\dag  \hat a_i \hat a_i ^\dag  \hat a_i \rangle
&=& n^2 + (2n+1)\sum\limits_{i \in \{ \eta\} } {\left| {C_{\{
\eta\} } }
\right|^2 }\nonumber\\
&=&n^2 + (2n+1)\frac{{M_1 }}{M}\nonumber\\
\Rightarrow  \Delta n^2 &=& \langle \hat a_i ^\dag  \hat a_i \hat a_i ^\dag
\hat a_i \rangle_i - \langle \hat a_i ^\dag  \hat a_i
\rangle_i^2\nonumber\\
&=& \frac{{M_1 }}{M}(1-\frac{{M_1 }}{M}).
\end{eqnarray}
For the off-diagonal terms,
\begin{eqnarray}
\langle \hat a_i ^\dag  \hat a_j \rangle  &=& \sum\limits_{\{
\eta_2 \ldots \eta_{M_1 } \} } {\frac{{C_{\{ i,\eta_2  \ldots
\eta_{M_1 } \} }^* C_{\{ j,\eta_2  \ldots \eta_{M_1 } \} } }}{{(n
+ 1)}}\langle {\hat a_i
\hat a_i ^\dag \rangle_i \langle \hat a_j \hat a_j ^\dag  } \rangle_j }\nonumber\\
&=& (n + 1)\sum\limits_{\{ \eta_2  \ldots \eta_{M_1 } \} } {C_{\{
i,\eta_2 \ldots \eta_{M_1 } \} }^* C_{\{ j,\eta_2  \ldots
\eta_{M_1 } \} } },\nonumber
\end{eqnarray}
where $ {\{ \eta_2  \ldots \eta_{M_1 } \} }$ is a set of distinct
integers chosen from $\{ 1,2, \ldots ,M\}$ excluding $i$ and $j$
and $\sum {_{\{ \eta_2  \ldots \eta_{M_1 } \} } } $ denotes the
sum over all combinatorial configurations. Because $C_{\{ \eta\}
}$ is determined by the energy cost of the corresponding state,
the difference between $ {C_{\{ i,\eta_2  \ldots \eta_{M_1 } \} }
} $ and $ {C_{\{ j,\eta_2 \ldots \eta_{M_1 } \} } }$ is estimated
to be of order $1/M_1$, and hence vanishes in the thermodynamic
limit. By considering symmetry and the normalization condition, we
find
\begin{eqnarray}
\langle \hat a_i ^\dag  \hat a_j \rangle  &=& (n +
1)\sum\limits_{\{ \eta_2 \ldots \eta_{M_1 } \} } {\left| {C_{\{
i,\eta_2
\ldots \eta_{M_1 } \} } } \right|^2 }\nonumber\\
 &=& (n+1)\frac{{{\textrm{\#
of terms in the sum }}\sum {{}_{\{ \eta_2  \ldots \eta_{M_1 } \}
}} }}{{{\textrm{\# of terms in the sum }}\sum {{}_{\{ \eta\} }}
}}\nonumber\\
 &=& (n+1)\frac{{\left( {\begin{array}{*{20}c}
   {M - 2}  \\
   {M_1  - 1}  \\
\end{array}} \right)}}{{\left( {\begin{array}{*{20}c}
   M  \\
   {M_1 }  \\
\end{array}} \right)}} = (n+1)\Delta n^2.
\end{eqnarray}

In the position representation, the density matrix is of the form
with all diagonal elements $x=n+M_1/M$ and all off-diagonal ones
$y=(n+1)\Delta n^2$ and can be diagonalized by Fourier
transformation to the momentum representation:
\begin{eqnarray}
 \hat \rho  &=& (x - y)I + y\sum\limits_{i,j} {\left| i \right\rangle \left\langle j \right|}  \nonumber  \\
&=& (x - y)\sum\limits_k {\left| k \right\rangle \left\langle k \right|}  + yM\left| {k = 0} \right\rangle \left\langle {k = 0} \right|  \nonumber \\
&=& [x + (M - 1)y]\left| {k = 0} \right\rangle \left\langle {k = 0} \right| + (x - y)\sum\limits_{k \ne 0} {\left| k \right\rangle \left\langle k \right|}  \nonumber  \\
\end{eqnarray}
Because $M$ is much larger than $1$, the $k=0$ state has much
larger occupation than all the other states, which are uniformly
occupied in a small fraction. The condensate fraction of the
system is given by $[x + (M - 1)y]/N$.

\section{Pseudo-spin model and Goldstone modes}

In the rf transition process, the system with $M$ sites can be
well described in the mean-field approach as a product of $M$
single-site states. Up to leading order in $J/U$, each single-site
state is represented within a truncated basis of four vectors, $|
n \rangle \otimes | 0 \rangle$, $| n+1 \rangle \otimes | 0
\rangle$, $| n \rangle \otimes | 1 \rangle$, and $| n+1 \rangle
\otimes | 1 \rangle$, where we use the notation
$|n_a+n_b,n_b\rangle$. As each number has only two possible values
in the truncated bases, we can map these states onto the states of
two spin-$1/2$ spins as $| \downarrow\rangle \otimes |
\downarrow_b \rangle$, $| \uparrow \rangle  \otimes |\downarrow_b
\rangle$, $| \downarrow \rangle \otimes |\uparrow_b \rangle$, and
$| \uparrow \rangle \otimes |\uparrow_b \rangle$ respectively.
Consistent with this mapping of the basis, any operator in the
Bose-Hubbard Hamiltonian of Eq. (\ref{eqn:2SBH}) can be replaced
by its corresponding spin operator as long as they have the same
matrix representation in the truncated space. Therefore we have
\begin{eqnarray}
\hat n_a  + \hat n_b  &\to& n + \frac{1}{2} + \hat S^z, \nonumber\\
\hat n_b  &\to& \frac{1}{2} + \hat S_b^z,\nonumber\\
\hat a &\to& \sqrt {n + 1/2 -\hat S_b^z } \hat S^-,\nonumber\\
\hat b^\dag  \hat a &\to& \sqrt {n + 1/2 + \hat S^z } \hat S_b^+,
\end{eqnarray}
where ${\bf{\hat{S}}}={\bf\hat{\sigma}}/2$, and ${\bf
\hat{\sigma}}$ is the vector of Pauli spin matrices. With these
substitutions, we obtain the pseudo-spin Hamiltonian for the
two-species Bose-Hubbard Hamiltonian of Eq. (\ref{eqn:2SBH}), up
to a constant:
\begin{eqnarray}
\hat H_{PS}  =  - J\sum\limits_{\left\langle {ij} \right\rangle } \sqrt {n +
1/2 - \hat S_{bi}^z } \sqrt {n + 1/2 - \hat S_{bj}^z
}\nonumber\\
\times\left( {\hat S_i^x \hat S_j^x  + \hat S_i^y \hat S_j^y } \right) +
\sum\limits_i {\left[ {(Un - \mu )\hat S_i^z  + \omega _b \hat S_{bi}^z
}\right]}.\label{eqn:HPS}
\end{eqnarray}
This Hamiltonian describes a system with two spins (${{\bf{\hat
S}}}$ and ${{\bf{\hat S_b}}}$) on each site, each of which
responds to an independent external ``magnetic field" in the
$z$-direction. The interaction between transverse components of
adjacent ${{\bf{\hat S}}}$ spins has a strength related to the
$z$- components of the $b$ spins on those sites. When
$S_{bi}^z=-1/2$, which corresponds to having no particles in the
$b$ hyperfine state, $\hat H_{PS}$ becomes the pseudospin
approximation to the single-species Bose-Hubbard
model~\cite{Barankov07}. The ground state of $\hat H_{PS}$ is the
same as Eq. (\ref{eqn:2SGS}), when written in the corresponding
spin basis.

The equivalence of the matrix representations of the Bose-Hubbard
Hamiltonian and the pseudo spin Hamiltonian imply that they also
have the same excitations in the truncated space. In fact, the
Goldstone excitations of the pseudo spin model, the spin waves,
are equivalent to those of the Bose-Hubbard model, which are small
density and phase distortion over a long length scale. To obtain
the dispersion relations of the spin waves, we introduce two
species of Holstein-Primakoff bosons ($A$ and $B$) to represent
small fluctuations of the corresponding spins around their
equilibrium value at zero temperature. Provided the fluctuations
are sufficiently small such that all terms higher than quadratic
can be neglected, the spin operators can be represented as $
{\bf{\hat S}} = (\frac{{\cos \theta }}{2}(\hat A + \hat A^\dag  )
+ \sin \theta (\frac{1}{2} - \hat A^\dag  \hat A),\frac{i}{2}(\hat
A + \hat A^\dag  ), - \frac{{\sin \theta }}{2}(\hat A + \hat
A^\dag  ) + \cos \theta (\frac{1}{2} - \hat A^\dag  \hat A))$ and
$ {\bf{\hat S}}_b  = ( - \frac{1}{2}(\hat B + \hat B^\dag
),\frac{i}{2}( - \hat B + \hat B^\dag  ), - \frac{1}{2} + \hat
B^\dag \hat B)$. After substituting these into the pseudo-spin
Hamiltonian of Eq. (\ref{eqn:HPS}) and keeping terms up to
quadratic order in the Holstein-Primakoff bosons, we Fourier
transform the resulting Hamiltonian into momentum space and
perform a Bogoliubov transformation on the ${A}$ bosons into
${\alpha}$ bosons ($ \hat A_p = u_p \hat \alpha _p  + v_p \hat
\alpha _{ - p}^\dag $) to diagonalize the $A$ terms. Thus we
arrive at the diagonalized Hamiltonian:
\begin{equation}
\hat H_{PS}  = \sum\limits_{p =  {-\pi/l} }^{\pi/l}  {\varepsilon _{p} \alpha
_{p}^\dag  \alpha _{p}  +  \omega _B \hat B_{p} ^\dag \hat B_{p} },
\label{eqn:HHP}
\end{equation}
where $p$ is the lattice momentum and $l$ the lattice spacing. The
$B$ terms correspond to the creation of a $B$ boson with momentum
$p$, while the $\alpha_p$ operator creates a Goldstone excitation
in the condensate with momentum $p$. The excitation energy of the
${\alpha}$ excitations is $\varepsilon _p = ZJ(n + 1)\sqrt {(1 -
I_p \cos ^2 \theta )(1 - I_p )}$, where $ I_p = \sum\limits_{i =
1}^d {d^{ - 1} \cos p_i l}$ for $d$ dimension. For small $p$
($pl\ll 1$), $\varepsilon _p = J(n+1)\sqrt {Z\sin ^2 \theta (pl)^2
+ \cos ^2 \theta (pl)^4 }$, which means that in the deep
condensate regime ($ \sin \theta  \gg pl $) the low-energy
excitations are wave-like (energy $\propto p$), while near the
Mott-insulator boundary ($ \sin \theta \ll pl $) they are
particle-like (energy $\propto p^2$). The excitation energy of of
the ${B}$ particles is $ \omega _B = \omega _{ba} + (1/8)ZJ\sin
\theta$, which is independent of $p$. This excitation is gapped
(reflecting the fact that the $b$ bosons are in their
Mott-insulator phase) and therefore should not be present in the
initial state in rf transitions for temperatures $kT \ll
\hbar\omega_{ba}$, which we assume throughout.

\begin{figure}[t]
\centering
    \includegraphics[width=7cm]{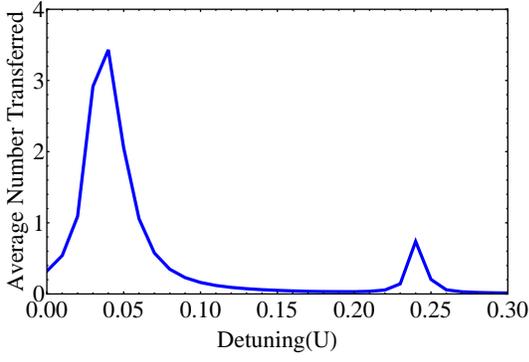}
  \caption{Average fraction transferred of a 6-boson 4-site entangled state
  driven by the external rf field. The parameters are $U_a=U_{ab}=1$,
  $ZJ=0.05$, $\mu=1.5$, and $U_b=\omega_{ba}=5$. We can see the spectrum
  has 2 peaks with positive shifts of order $J$.
  }
    \label{fig:f08}
\end{figure}

\begin{figure}[t]
\centering
    \includegraphics[width=7cm]{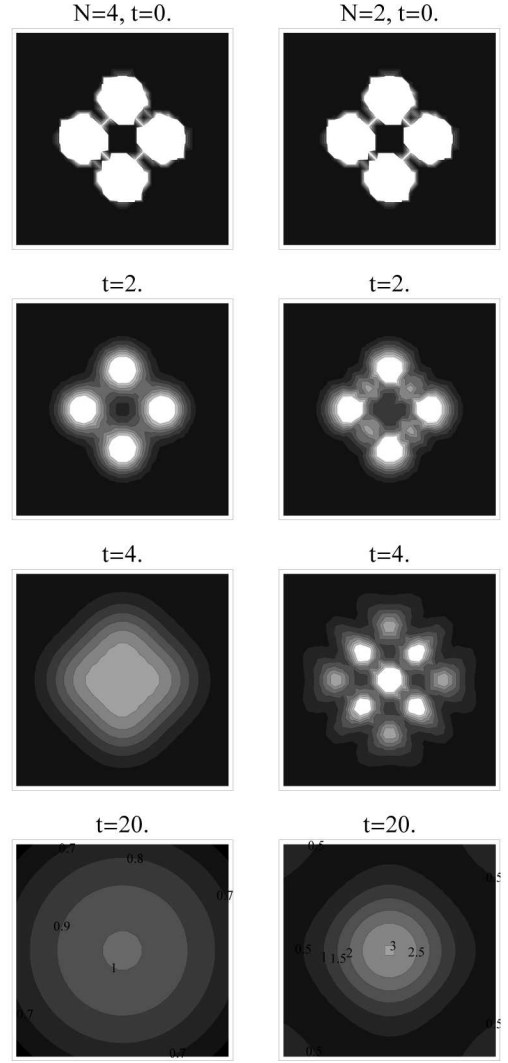}
  \caption{Matter wave interference patterns (per particle contribution).
  The left column is the 4-particle 4-site Mott-insulator, and the
  right one is the 2-particle 4-site condensate. The four rows from top to bottom
  are at $t=0$, $2$, $4$, and $20$ time units($ml^2/2 \hbar$). All parts of the figure have the same spatial scale.
  The number in the last row is the relative value of the density.}
    \label{fig:f09}
\end{figure}

 Turning to the role of the Goldstone modes in rf transitions, the form of the
 rf perturbation (given in the position basis of the physical particles as
 $\sum\limits{}_i\gamma\hat{a}_i^\dag \hat{b}_i+h.c.$) is given in the Goldstone
 basis by
\begin{eqnarray}
H_{rf} =  &-& \sqrt N \gamma \hat B_0 ^\dag
 + \frac{{\gamma \sin \theta }}{{4\sqrt {\bar n} }}\sum\limits_p {(v_p  + u_p )
 \hat B_p ^\dag  (\alpha _p  + \alpha _{ - p}^\dag  )}\nonumber\\
 &+& \rm {H.c.}, \label{eqn:RFG}
\end{eqnarray}
where $N$ is the total number of particles and ${\bar n}=n+\cos ^2
(\theta /2)$ is the number density at zero temperature. The first
 term creates a $B$ boson uniformly in space without any distortions in density or phase (which are represented by $\alpha$
bosons). The second term creates a $B$ boson with momentum $p$ accompanied by
an $\alpha$ boson with momentum $-p$ or annihilation of an $\alpha$ boson with
momentum $p$.

For the 3D system at zero temperature, using the above Goldstone
representation of the rf perturbation in Fermi's golden rule [(Eq.
\ref{eqn:FGR})], we find the transition rate due to the first term
of Eq. (\ref{eqn:RFG}) is $I^{(1)}(\omega) \propto \delta
(\omega-\omega_B)$. This delta peak corresponds to the first peak
we obtained in Sec. IIIA, although slightly shifted given the
slightly different mean-field energy estimates. The transition
rate due to the second term is $I^{(2)}(\omega) \propto
\sin^2\theta |\omega-\omega_B|^3$ which is so small compared to
the delta function near $\omega_B$ that the effect of $I^{(2)}$ on
the contrast of $I^{(1)}$  can be ignored. We expect similar
argument to hold for the second peak obtained in
Sec.~\ref{sec:IIIA}. Therefore, we find that excitation of
Goldstone modes by a weak rf field will not obscure the two-peak
signature at zero temperature. At low temperatures of order $kT\ll
ZJ$, the initial state still contains zero $B$ bosons but does
contain $\alpha$ bosons (corresponding to Goldstone modes) with a
bosonic thermal distribution. Taking these thermally-excited
$\alpha$ bosons into consideration in the rf signal, we find that
$I^{(1)}$ is still a delta peak, while $I^{(2)}$ becomes a finite
function of $|\omega-\omega_B|$ which does not obscure the peak
$I^{(1)}$. For higher temperatures of order $ZJ$ (when the average
number of $\alpha$ bosons per site is of order $1$ or more), the
higher order terms in $H_{rf}$ cannot be ignored.

\section{Time-evolution of a six-boson four-site system}

The dynamics of finite bosons in a small number of lattice sites
can be numerically solved without using the mean-field approach,
and the results reflect the physical signatures predicted in our
theoretical analysis. Here we consider a system of six bosons on
four sites located on a ring. The ground state is given by
\begin{eqnarray}
&{}&\frac{1}{{2\sqrt 2 }}( \left| 1 \right\rangle \left| 1 \right\rangle \left|
2 \right\rangle \left| 2 \right\rangle  + \left| 1 \right\rangle \left| 2
\right\rangle \left| 2 \right\rangle \left| 1 \right\rangle  + \left| 2
\right\rangle \left| 2 \right\rangle \left| 1 \right\rangle \left| 1
\right\rangle \nonumber\\
&{}&+ \left| 2 \right\rangle \left| 1 \right\rangle \left| 1 \right\rangle
\left| 2 \right\rangle) + \frac{1}{2}\left( \left| 1 \right\rangle \left| 2
\right\rangle \left| 1 \right\rangle \left| 2 \right\rangle  + \left| 2
\right\rangle \left| 1 \right\rangle
\left| 2 \right\rangle \left| 1 \right\rangle \right)\nonumber\\
&{}&+{\rm O}(J/U),
\end{eqnarray}
where $\left| {n_1 } \right\rangle \left| {n_2 } \right\rangle
\left| {n_3 } \right\rangle \left| {n_4 } \right\rangle$ means the
state with the $i$th site occupied by $n_i$ bosons. The ground
state has nonzero single-site number fluctuations, which
differentiate the condensate from the Mott-insulator. Thus the
system is expected to have an rf spectrum similar to that of a
many-particle condensate. We use the Bose-Hubbard Hamiltonian,
plus time-dependent terms which represent the interaction with the
rf field, to numerically time evolve the ground state, and compute
the average number of bosons transferred from the $a$ state to the
$b$ state (Fig.~\ref{fig:f08}). The result shows two peaks with
shifts of order $J$, as we obtained in Sec.~\ref{sec:II} through
the mean-field approach.

To show matter-wave interference, we calculate the density
contribution per particle of a two-particle four-site system
(condensate) and that of a four-particle four-site system (the
Mott-insulator) in free expansion(Fig.~\ref{fig:f09}). Both cases
have the same initial density profile but have very different
interference patterns in expansion. The condensate shows more
interference peaks and has a much higher density in the center at
long times than the Mott-insulator does. This difference is
similar to the difference between the phase-coherent state
(superfluid) and the phase-incoherent state (the Mott-insulator)
in the experiment of Ref.~\cite{Greiner02}.

\end{document}